\documentclass[%
reprint,
superscriptaddress,
amsmath,amssymb,
]{revtex4-2}

\usepackage{xcolor}

\definecolor{codegreen}{rgb}{0,0.6,0}
\definecolor{codegray}{rgb}{0.5,0.5,0.5}
\definecolor{codepurple}{rgb}{0.58,0,0.82}
\definecolor{backcolour}{rgb}{0.95,0.95,0.92}
\definecolor{urlblue}{HTML}{007bff}

\usepackage{graphicx}% Include figure files
\usepackage{dcolumn}% Align table columns on decimal point
\usepackage{bm}% bold math
\usepackage{orcidlink}
\usepackage{hyperref}% add hypertext capabilities
\usepackage{cleveref}
\usepackage{microtype}

\newcommand{\Tr}{\ensuremath{\textup{Tr}}}
\renewcommand{\bf}[1]{\ensuremath{\textbf{#1}}}

\hypersetup{
    colorlinks = true,
    linkcolor  = urlblue,
    citecolor  = urlblue,
    urlcolor   = urlblue,
}

\usepackage[mathlines]{lineno}% Enable numbering of text and display math

\begin{document}

% \preprint{}

\title{An Exact Chiral Amorphous Spin Liquid}

\author{G. Cassella \orcidlink{0000-0003-4506-5791}}
\thanks{These three authors contributed equally, and names are ordered alphabetically.}
\affiliation{\small Blackett Laboratory, Imperial College London, London SW7 2AZ, United Kingdom}

\author{P. d'Ornellas \orcidlink{0000-0002-2349-0044}}
\thanks{These three authors contributed equally, and names are ordered alphabetically.}
\affiliation{\small Blackett Laboratory, Imperial College London, London SW7 2AZ, United Kingdom}

\author{T. Hodson \orcidlink{0000-0002-4121-4772}}
\thanks{These three authors contributed equally, and names are ordered alphabetically.}
\affiliation{\small Blackett Laboratory, Imperial College London, London SW7 2AZ, United Kingdom}

\author{W. M. H. Natori \orcidlink{0000-0002-0740-2956}}
\affiliation{Institut Laue-Langevin, BP 156, 41 Avenue des Martyrs, 38042 Grenoble Cedex 9, France}

\author{J. Knolle \orcidlink{0000-0002-0956-2419}}
\affiliation{\small Blackett Laboratory, Imperial College London, London SW7 2AZ, United Kingdom}
\affiliation{Department of Physics TQM, Technische Universit{\"a}t M{\"u}nchen, James-Franck-Stra{\ss}e 1, D-85748 Garching, Germany}
\affiliation{Munich Center for Quantum Science and Technology (MCQST), 80799 Munich, Germany}

\date{\today}

\begin{abstract}
Topological insulator phases of non-interacting particles have been generalized from periodic crystals to amorphous lattices, which raises the question whether topologically ordered quantum many-body phases may similarly exist in amorphous systems? Here we construct a soluble chiral amorphous quantum spin liquid by extending the Kitaev honeycomb model to random lattices with fixed coordination number three. The model retains its exact solubility but the presence of plaquettes with an odd number of sides leads to a spontaneous breaking of time reversal symmetry. We unearth a rich phase diagram displaying Abelian as well as a non-Abelian quantum spin liquid phases with a remarkably simple ground state flux pattern. Furthermore, we show that the system undergoes a finite-temperature phase transition to a conducting thermal metal state and discuss possible experimental realisations. 
\end{abstract}

\maketitle

\section{Introduction}

Amorphous materials are condensed matter systems characterised by short-range regularities, and an absence of long-range crystalline order as studied early on for amorphous semiconductors~\cite{Yonezawa1983,zallen2008physics}. The bonds of a wide range of covalent compounds can enforce local constraints around each ion, e.g.~a fixed coordination number $z$, which has enabled the prediction of energy gaps even in lattices without translational symmetry~\cite{Weaire1976,gaskell1979structure}, the most famous example being amorphous Ge and Si with $z=4$~\cite{Weaire1971,betteridge1973possible}. Recently, following the discovery of topological insulators (TIs), it has been shown that similar phases can exist in amorphous systems characterized by protected edge states and topological bulk invariants~\cite{mitchellAmorphousTopologicalInsulators2018,agarwala2019topological,marsalTopologicalWeaireThorpeModels2020,costa2019toward,agarwala2020higher,spring2021amorphous,corbae2019evidence}. However, research on electronic systems has been mostly focused on non-interacting systems with few exceptions, for example, to account for the observation of superconductivity~\cite{buckel1954einfluss,mcmillan1981electron,meisel1981eliashberg,bergmann1976amorphous,mannaNoncrystallineTopologicalSuperconductors2022} in amorphous materials or very recently to understand the effect of strong electron repulsion in TIs~\cite{kim2022fractionalization}.     

Magnetism in amorphous systems has been investigated since the 1960s, mostly through the adaptation of theoretical tools developed for disordered systems \cite{aharony1975critical,Petrakovski1981,kaneyoshi1992introduction,Kaneyoshi2018} and with numerical methods~\cite{fahnle1984monte,plascak2000ising}. Research has focused on classical Heisenberg and Ising models, which are able to describe ferromagnetic, disordered antiferromagnetic and widely observed spin glass behaviour~\cite{coey1978amorphous}. However, the role of spin-anisotropic interactions and quantum effects in amorphous magnets has not been addressed. It is an open question whether frustrated magnetic interactions on amorphous lattices can give rise to genuine quantum phases, i.e.~to long-range entangled quantum spin liquids (QSL)~\cite{Anderson1973,Knolle2019,Savary2016,Lacroix2011}. 

The combination of a fixed local coordination number in conjunction with magnetic frustration generated by bond-anisotropic Ising exchanges can lead to stable QSL phases. The seminal Kitaev model on the honeycomb lattice~\cite{kitaevAnyonsExactlySolved2006} provides an exactly solvable model whose ground state is a QSL characterized by a static $\mathbb Z_2$ gauge field and Majorana fermion excitations. Several instances of Kitaev candidate materials have been synthesized in the last decade \cite{Jackeli2009,HerrmannsAnRev2018,Winter2017,TrebstPhysRep2022,Takagi2019} following the suggestion that heavy-ion Mott insulators formed by edge-sharing octahedra may realize dominant Kitaev interactions \cite{Jackeli2009}. In particular, recently it has been shown that the Kitaev material Li\textsubscript{2}IrO\textsubscript{3} can be created with an amorphous structure \cite{lee_lithiair_2019}. In fact, with sufficiently fast cooling, any crystalline material can be made amorphous \cite{zallen2008physics, turnbull_1969}, opening the possibility for exploring a wide variety of non-crystalline Kitaev materials.

It is by now well known that the Kitaev model on any three-coordinated ($z=3$) graph has conserved plaquette operators and local symmetries~\cite{Baskaran2007,Baskaran2008} which allow for a mapping onto an effective free Majorana fermion problem in a background of static $\mathbb Z_2$ fluxes~\cite{Nussinov2009,OBrienPRB2016,yaoExactChiralSpin2007,hermanns2015weyl}. However, in general this neither  means that any $z=3$ lattice Kitaev model can be straightforwardly constructed, nor that the QSL properties are obvious. Several obstacles  remain. First, the labelling of bonds necessary to create a soluble Hamiltonian can be an NP-complete problem. Second, once the Majorana system has been constructed, determining the ground state out of the exponentially large number of $\mathbb Z_2$ flux sectors is generically hard, since Lieb's theorem -- which defines the ground state flux configuration for the honeycomb --  is not applicable for most lattices. Previous studies have relied on translation and reflection symmetries to reduce the number of sectors that must be checked \cite{yaoExactChiralSpin2007,eschmann2019thermodynamics,Peri2020}, which cannot be done in an amorphous system. Third, once the ground state flux sector is found, it needs to be determined whether lattice disorder induces a gapless phase~\cite{selfThermallyInducedMetallic2019,Laumann2012,lahtinenTopologicalLiquidNucleation2012,selfThermallyInducedMetallic2019} or whether the fermionic spectrum is gapped, possibly with non-trivial topology~\cite{yaoExactChiralSpin2007}.

In this article we study the Kitaev model on amorphous lattices and establish it as an example of a topologically ordered amorphous QSL phase. Concentrating on random networks generated via Voronoi tessellation \cite{mitchellAmorphousTopologicalInsulators2018,marsalTopologicalWeaireThorpeModels2020} with $z=3$, we show how to colour the bonds consistently. We find that the presence of plaquettes with an odd-number of sites lead to a chiral QSL with spontaneously broken time-reversal symmetry (TRS)~\cite{yaoExactChiralSpin2007,Chua2011,ChuaPRB2011,Fiete2012,Natori2016,Wu2009, WangHaoranPRB2021}. We establish via extensive numerics that the ground state $\mathbb Z_2$ flux sector follows a remarkably simple counting rule consistent with Lieb's theorem \cite{lieb_flux_1994}. We map out the phase diagram of the model and show that the chiral phase around the symmetric point is gapped and characterized by a quantized local Chern number $\nu$~\cite{dornellas2022, mitchellAmorphousTopologicalInsulators2018} as well as protected chiral Majorana edge modes. Finally, we discuss the effect of additional bond disorder and comment on the role of finite temperature fluctuations, showing that the proliferation of flux excitations leads to an Anderson transition to a thermal metal phase \cite{Laumann2012,lahtinenTopologicalLiquidNucleation2012,selfThermallyInducedMetallic2019}. \par
\begin{figure*}[t]
    \centering
    \includegraphics[width=0.95\textwidth]{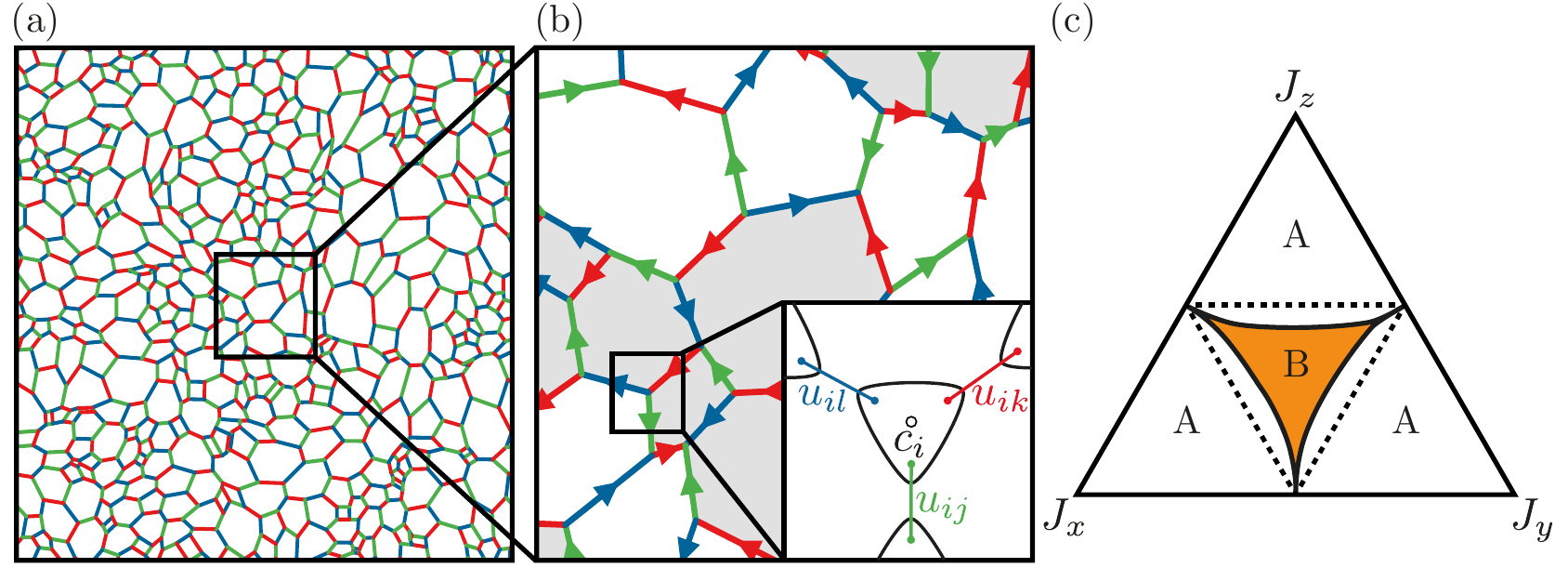}
    \caption{ Construction details for the amorphous lattice model. \textbf{(a)} Amorphous lattice generated via Voronoi tessellation of a uniformly distributed random point set on the unit square. Periodic boundary conditions are imposed by tiling the unit square before Voronoi tessellation. \textbf{(b)} Magnified portion of the amorphous lattice. Arrows from site $j$ to site $k$ indicate  direction where the bond variable $u_{jk} = 1$. An arbitrary flux sector is shown, where shaded plaquettes have $\mathbb Z_2$ flux flipped with respect to the ground state. Colours correspond to a valid assignment of the bond colourings, $\alpha_{jk}$. The inset demonstrates the Majorana construction on a tri-coordinate motif, which allows for the exact solution of the model. \textbf{(c)} Ternary phase diagram of the amorphous Kitaev model with varying exchange coupling. The isotropic regime $|J_x| \approx |J_y| \approx |J_z|$ (B), exhibits a topologically non-trivial chiral QSL ground state with Chern number $\nu=\pm1$. The fermion gap of the ground state flux sector closes at the phase boundary (solid black lines), and a transition occurs to a $\nu=0$ phase (A) for anisotropic couplings. The phase boundary was obtained by averaging over 20 amorphous lattice realisations with $\sim400$ sites. Dotted black lines indicate the corresponding phase boundaries in the honeycomb model.}
    \label{fig:example_lattice}
\end{figure*}

\section{Methods}
We start with a brief review of the Kitaev model on the honeycomb lattice \cite{kitaevAnyonsExactlySolved2006} before generalising it to amorphous systems. A spin-1/2 is placed on every vertex and each bond is labelled by an index $\alpha \in \{ x, y, z\}$. The bonds are arranged such that each vertex connects to exactly one bond of each type. The Hamiltonian is given by
\begin{equation}
    \label{eqn:kitham}
    \mathcal{H} = - \sum_{\langle j,k\rangle_\alpha} J^{\alpha}\sigma_j^{\alpha}\sigma_k^{\alpha},
\end{equation}
where $\sigma^\alpha_j$ is a Pauli matrix acting on site $j$, \(\langle j,k\rangle_\alpha\) is a pair of nearest-neighbour indices connected by an $\alpha$-bond with exchange coupling $J^\alpha$. For each plaquette of the lattice, we define a flux operator $ W_p = \prod \sigma_j^{\alpha}\sigma_k^{\alpha}$, where the product runs clockwise over the bonds around the plaquette. These commute with one another and the Hamiltonian, so correspond to an extensive number of conserved quantities. This allows us to split the Hilbert space according to the eigenvalues $\phi_p = \pm 1$ ($\pm i$ for odd plaquettes) of $\{W_p\}$.

The Hamiltonian in \cref{eqn:kitham} can be exactly solved by transforming to a Majorana fermion representation~\cite{kitaevAnyonsExactlySolved2006}, see fig.~\ref{fig:example_lattice}. Each spin is represented with four Majorana operators, $\sigma_i^\alpha = i b_i^\alpha c_i$. We define a set of conserved bond operators $\hat u_{jk} = ib_j^{\alpha}b_k^{\alpha}$. As with the $W_p$ operators, we may partition the Majorana Hilbert space according to the eigenvalues of these operators, $u_{jk}=\pm 1$. For a given choice of these bond variables, \cref{eqn:kitham} reduces to a quadratic Majorana Hamiltonian
\begin{equation}\label{eqn:majorana_hamiltonian}
    \mathcal{H} = \frac{i}{4}\sum_{ j,k }A_{jk} c_j c_k,
\end{equation}
where $A_{jk}=2J^{\alpha}u_{jk}$. 

The matrix $iA$ determines properties of the fermionic degrees of freedom for a given flux configuration $\{u_{jk} \}$. The spectrum is obtained by rotating to a new Majorana basis consisting of pairs of operators $\tilde c_j', \tilde c_j''$, defined by a matrix $\tilde c_j = R_{jk}c_k $ containing the fermionic eigenstates. The Hamiltonian takes the form $\mathcal{H} = \sum_{ j } \varepsilon_j\, i \tilde c_j' \tilde c_j''$, and in what follows we refer to fermionic properties of the system as those determined by $iA$ in a fixed flux sector.

The Kitaev Hamiltonian remains exactly solvable on any lattice in which no site connects to more than one bond of the same type \cite{Nussinov2009}. Thus, we shall restrict our investigation to lattices in which every vertex has coordination number $z \leq 3$. Here we generate such lattices using Voronoi tessellation~\cite{florescu_designer_2009}. Once a lattice has been generated, the bonds must be labelled in such a way that no vertex touches multiple edges of the same type, which we refer to as a three-edge colouring. The problem of finding such a colouring is equivalent to the classical problem of four-colouring the faces, which is always solvable on a planar graph \cite{Tait1880, appelEveryPlanarMap1989a} but can  take up to seven colours on the torus~\cite{ringel_solution_1968}. In practice, we reduce the colouring to a Boolean satisfiability problem~\cite{Karp1972} with details described in the supplemental material. One example of a coloured amorphous lattice is shown in 
fig.~\ref{fig:example_lattice}(a). 

Once the lattice and colouring has been found, the amorphous Hamiltonian is diagonalised using the same procedure as for the honeycomb model. Note that the Majorana system is only strictly equivalent to the initial spin system after a parity projection~\cite{pedrocchiPhysicalSolutionsKitaev2011, Yao2009}, details of which for the amorphous case are described in the supplemental material. Nevertheless, one can still use \cref{eqn:majorana_hamiltonian} to evaluate the expectation values of operators that conserve $\hat u_{jk}$ in the thermodynamic limit \cite{zschocke2015physical,knolle_dynamics_2016}. The ground state energy of a given flux sector is the sum of the negative eigenvalues of $iA/4$ in \cref{eqn:majorana_hamiltonian}, and excitation energies are given by the positive eigenvalues. 

\section{Results}
We first investigate which flux patterns minimize the ground state energy on the amorphous lattice. When represented in the Majorana Hilbert space, flux operators $ W_p = \prod \sigma_j^{\alpha}\sigma_k^{\alpha}$ correspond to ordered products of link variables $\hat u_{jk}$, and their eigenvalues describe the $\mathbb Z_2$ flux through each plaquette,
\begin{equation} \label{eqn:flux_definition}
    \phi_p = \prod_{(j,k) \in \partial p} -iu_{jk},
\end{equation}
where the product is taken over the $u_{jk}$ values going clockwise around the border $\partial p$ of each plaquette. We refer to a particular choice of a set of $\{ \phi_p\}$ as a flux sector.\par
The spin Hamiltonian is real, thus it has TRS. However, the flux $\phi_p$ through any plaquette with an odd number of sides has imaginary eigenvalues $\pm i$. Thus, states with a fixed flux sector spontaneously break TRS, which in the context of crystalline Kitaev models was first described by Yao and Kivelson~\cite{Yao2011}. All flux sectors come in degenerate pairs, where time reversal is equivalent to inverting the flux through every odd plaquette~\cite{yaoExactChiralSpin2007, Peri2020}.

For a system with $n_p$ plaquettes in periodic boundaries, there are $2^{n_p-1}$ possible flux sectors, and in general it is a nontrivial task to determine which pair of flux sectors has the lowest energy. On the honeycomb lattice, the ground state was shown by Lieb to be flux free, $\phi_p=+1$ \cite{lieb_flux_1994}, however no such proof exists for amorphous lattices, since all lattice symmetries are broken. 

To numerically determine the ground state flux sector, we first test a large number of finite size lattices ($\sim 25,000$ lattices with 16 plaquettes), directly enumerating all possible flux configurations to find the lowest energy. In practice, care must be taken to account for finite size effects, as well as to ensure that the results hold as system size is increased -- detailed in the supplemental material. Remarkably, we find that the energy is always minimised by setting the flux through each plaquette $p$ to 
\begin{align} \label{eqn:gnd_flux}
    \phi^{\textup{g.s.}}_p = -(\pm i)^{n_{\textup{sides}}},
\end{align}
where $n_{\textup{sides}}$ is the number of edges that form the plaquette and the global choice of the sign of $i$ gives the two TRS-degenerate ground state flux sectors. The conjecture is consistent with results found on other regular lattices for which Lieb's theorem is not applicable~\cite{OBrienPRB2016}. Having identified the ground state, any other flux sector can be characterized by the configuration of vortices, i.e.~by the plaquettes whose flux is flipped with respect to $\left\{ \phi_p^{\textup{g.s.}} \right\}$.

The ground state phase diagram can then be determined by varying the strength of each bond type, $J^\alpha$ while remaining in the ground state flux sector, and we numerically calculate the ternary phase diagram shown in \cref{fig:example_lattice} (c). The diagram contains two distinct phases: close to the corners of the triangle, e.g.~$|J^z| \gg |J^x|, |J^y|$, the (A) phase is equivalent to the toric code on an amorphous lattice \cite{kitaev_fault-tolerant_2003}. The phase has a fermionic gap and supports Abelian excitations. Around the isotropic point $J^x = J^y = J^z$, the (B) phase is also gapped in contrast to the honeycomb case as a consequence of TRS breaking from the finite density of odd plaquettes. All lattices studied in this work were generated from a voronoi lattice with completely random seed points, and so had on average equal proportions of odd and even plaquettes. We will confirm below that the (B) phase is indeed a {\it chiral spin liquid}. 

As the values of $J^\alpha$ are varied, the fermionic gap closes at the boundary between the two phases. In the honeycomb model, the phase boundaries are located on the straight lines $|J^\alpha| = |J^\beta| + |J^\gamma|$, for any permutation of $\alpha, \beta, \gamma \in \{x,y,z\}$. We find that on the amorphous lattice these boundaries exhibit an inward curvature similar to honeycomb Kitaev models with flux \cite{Nasu_Thermal_2015} or bond \cite{knolle_dynamics_2016} disorder.

Note, the presence of the gapped B-phase is non-trivial and related to our choice of homogeneous couplings for each colour of the bonds. In the supplemental material we study the robustness of the B phase with respect to bond disorder, e.g.~a bond-length dependence of the interaction strength. In general one might expect disorder to lead to a gap closing, however we find that the gap is reduced but remains robust up to sizeable bond disorder.
\begin{figure}
    \centering
    \includegraphics[width=0.95\columnwidth]{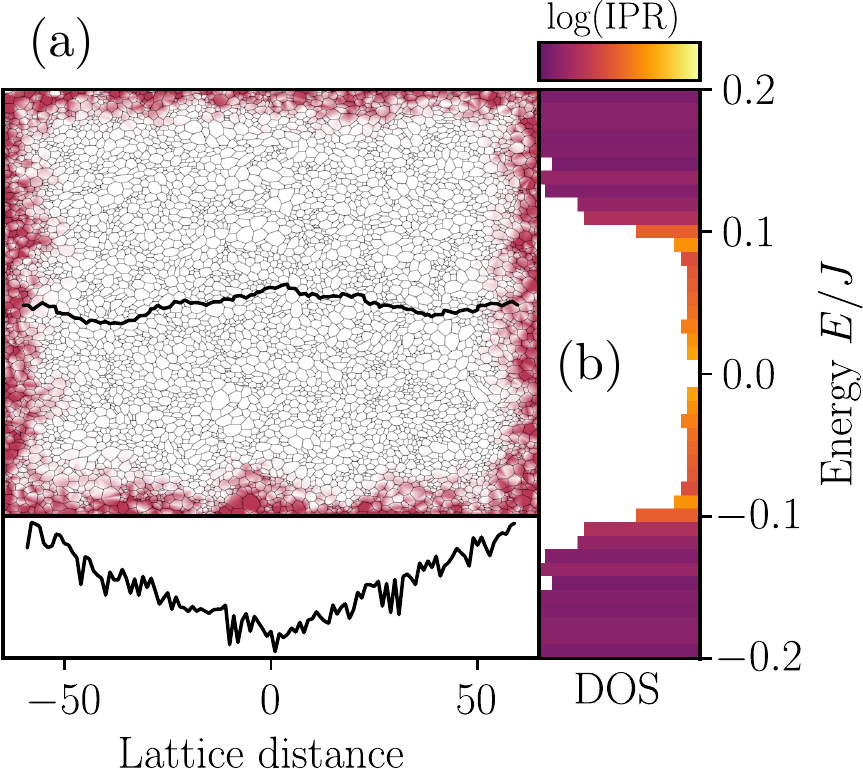}
    \caption{ Ground-state flux sector wavefunctions and spectrum. \textbf{(a)} In-gap fermionic wavefunction drawn from the ground state flux sector in open boundary conditions, showing a topological edge mode. The number density for this state along a line of lattice sites spanning the system (black line) is shown in the bottom subfigure on a logarithmic scale, demonstrating the characteristic exponential decay of topological edge modes with distance from the edge. \textbf{(b)} Ground-state flux sector fermionic density of states in open boundary conditions, coloured by inverse participation ratio. The increased inverse participation ratio of the in-gap states signifies their localisation to the edges of the system.}
    \label{fig:edge_modes}
\end{figure}

A fundamental tool for understanding the distinction between the two phases is the Chern number. The original definition relies on momentum space, and so cannot be used here, where the system lacks any translational symmetry. However, recently methods have been developed for evaluating a real-space analogue of the Chern number \cite{bianco_mapping_2011,Hastings_Almost_2010}. Here we shall use a slight modification of Kitaev's definition \cite{kitaevAnyonsExactlySolved2006, dornellas2022, mitchellAmorphousTopologicalInsulators2018}. For a choice of flux sector, we calculate the projector $P$ onto the negative energy eigenstates of the matrix $iA$ defined in \cref{eqn:majorana_hamiltonian}. The local Chern number around a point $\bf R$ in the bulk is given by 
\begin{align}
    \nu (\bf R) = 4\pi \textup{Im}\; \Tr_{\textup{Bulk}} 
    \left ( 
    P\theta_{R_x} P \theta_{R_y} P
    \right ),
\end{align}
where $\theta_{R_x}$ is a step function in the $x$-direction, with the step located at $x = R_x$, $\theta_{R_y}$ is defined analogously. The trace is taken over a region around $\bf R$ in the bulk of the material, where care must be taken not to include any points close to the edges. Provided that the point $\bf R$ is sufficiently far from the edges, this quantity will be very close to quantised to the Chern number.

Using this local Chern marker, we determine that the (A) phase has Chern number $\nu = 0$, whereas the two TRS-degenerate ground state flux sectors in the (B) phase have Chern number $\nu = \pm 1$ respectively. In closed boundaries, this leads to the appearance of gap-crossing protected edge modes, in accordance with the bulk-boundary correspondence \cite{qi_general_2006}, an example is shown in \cref{fig:edge_modes}. The edge modes are exponentially localised to the boundary of the system, and can be qualitatively distinguished from bulk states by their large inverse participation ratio,
\begin{equation}
    \textup{IPR} = \int d^2r|\psi(\mathbf{r})|^4,
\end{equation} 
where $\psi(\mathbf{r})$ denotes an eigenmode of the free Majorana Hamiltonian derived in \cref{eqn:majorana_hamiltonian}.
Finally, we note that the closing of the fermionic gap on the boundary between the two phases is necessary in order to transition between states with different Chern numbers.

{\it Anderson Transition to a Thermal Metal ---} 
Having understood the spontaneous formation of a chiral amorphous QSL ground state, we are now in a position to discuss the finite temperature behavior of the model. In general, an Ising-like thermal phase transition into the chiral QSL phase is expected akin to the one observed for the Yao-Kivelson model~\cite{nasu2015thermodynamics} but a full Monte-Carlo sampling, which is further complicated by the inherent disorder in the amorphous lattice, is beyond the scope of this letter. Nevertheless, the main effect of increasing temperature is the proliferation of fluxes which allow us to gain a qualitative understanding of the finite temperature behavior~\cite{Nasu_Thermal_2015}.

On the honeycomb Kitaev model with explicit TRS breaking, Majorana zero modes bind to fluxes forming Ising non-Abelian anyons \cite{Beenakker2013}. Their pairwise interaction decays exponentially with separation~\cite{Laumann2012,Lahtinen_2011,lahtinenTopologicalLiquidNucleation2012}. As temperature is increased, the proliferation of vortices in the system produces a finite density of anyons and their hybridization leads to an Anderson transition to a macroscopically degenerate state known as a \emph{thermal metal phase}~\cite{Laumann2012, selfThermallyInducedMetallic2019, Chalker_thermal_2001}. This exotic phase has two key signatures. Firstly, the metallic phase is defined by a closing of the fermion gap -- that is, it is driven by vortex configurations with a gapless fermionic spectrum. Secondly, we expect the density of states in a thermal metal to diverge logarithmically with energy and display characteristic low energy oscillations predicted by random matrix theory \cite{bocquet_disordered_2000, selfThermallyInducedMetallic2019}. In the supplemental material we present numerical evidence showing that all of the above features carry over to the amorphous QSL with spontaneous TRS breaking, giving strong evidence for the transition to the thermal metal phase.

\section{Discussion}

We have studied an extension of the Kitaev honeycomb model to amorphous lattices with coordination number $z= 3$. We found that it is able to support two quantum spin liquid phases that can be distinguished using a real-space generalisation of the Chern number. The presence of odd-sided plaquettes results in a spontaneous breaking of TRS, and the emergence of a chiral spin liquid phase. Furthermore we found evidence that the amorphous system undergoes an Anderson transition to a thermal metal phase, driven by the proliferation of vortices with increasing temperature. 
Our exactly soluble chiral QSL provides a first example of a topologically quantum many-body phase in amorphous magnets, which raises a number of questions for future research. 

First, a numerically challenging task would be a study of the full finite temperature phase diagram via Monte-Carlo sampling and possible violations of the Harris criterion for the Ising transition stemming from the inherent lattice disorder~\cite{barghathi2014phase,schrauth2018two,schrauth2018violation}. Second, it would be worthwhile to search for experimental realisations of amorphous Kitaev materials, which can possibly be created from crystalline ones using standards method of repeated liquifying and fast cooling cycles~\cite{Weaire1976,Petrakovski1981,Kaneyoshi2018}. The putative QSL behavior of the intercalated Kitaev compound H$_3$LiIr$_2$O$_6$~\cite{kitagawa2018spin,knolle2019bond} could possibly be related to amorphous lattice disorder. Moreover, metal organic frameworks are promising platforms forming amorphous lattices~\cite{bennett2014amorphous} with recent proposals for realizing strong Kitaev interactions~\cite{yamada2017designing} as well as reports of QSL behavior~\cite{misumi2020quantum}. We expect that an experimental signature of a chiral amorphous QSL is a half-quantized thermal Hall effect similar to magnetic field induced behavior of honeycomb Kitaev materials~\cite{Kasahara2018,Yokoi2021,Yamashita2020,Bruin2022}. Alternatively, it could be characterized by local probes such as spin-polarized STM~\cite{Feldmeier2020,Konig2020,Udagawa2021} and the thermal metal phase displays characteristic longitudinal heat transport signatures \cite{Beenakker2013}. Third, it would be interesting to study the stability of the chiral amorphous Kitaev QSL with respect to perturbations~\cite{Rau2014,Chaloupka2010,Chaloupka2013,Chaloupka2015,Winter2016} and, importantly, to investigate whether QSL may exist for spin-isotropic Heisenberg models on amorphous lattices. 

Overall, there has been surprisingly little research on amorphous quantum many body phases albeit material candidates aplenty. We expect our exact chiral amorphous spin liquid to find many generalisation to realistic amorphous quantum magnets and beyond. 

\vspace{0.3cm}
{\it Acknowledgements --- } 
We thank Adolfo Grushin and Cecille Repellin for helpful discussions and collaboration on related work. 
JK acknowledges support via the Imperial-TUM flagship partnership. The research is part of the Munich Quantum Valley, which is supported by the Bavarian state government with funds from the Hightech Agenda Bayern Plus. This work was supported in part by the Engineering and Physical Sciences Research Council (EP/T51780X/1 GC, EP/R513052/1 TH, PD).

{\it Data Availability --- }
The data used to produce these plots are of limited availability, due to the ease with which they can be generated from the publicly available code and, in the case of the evidence for the ground state flux sector, the large file size. Access can be obtained by contacting the authors.

{\it Code Availability --- }
The cource code used to generate these results is publicly available online \cite{koala}.

% \nocite{Karp1972,imms-sat18, koala, knolleBondDisorderedSpinLiquid2019}

\bibliography{refs}% Produces the bibliography via BibTeX.
\appendix

\section{Lattice Generation} \label{apx:lattice_construction}

\begin{figure*}[t]
    \centering
    \includegraphics[width=\textwidth]{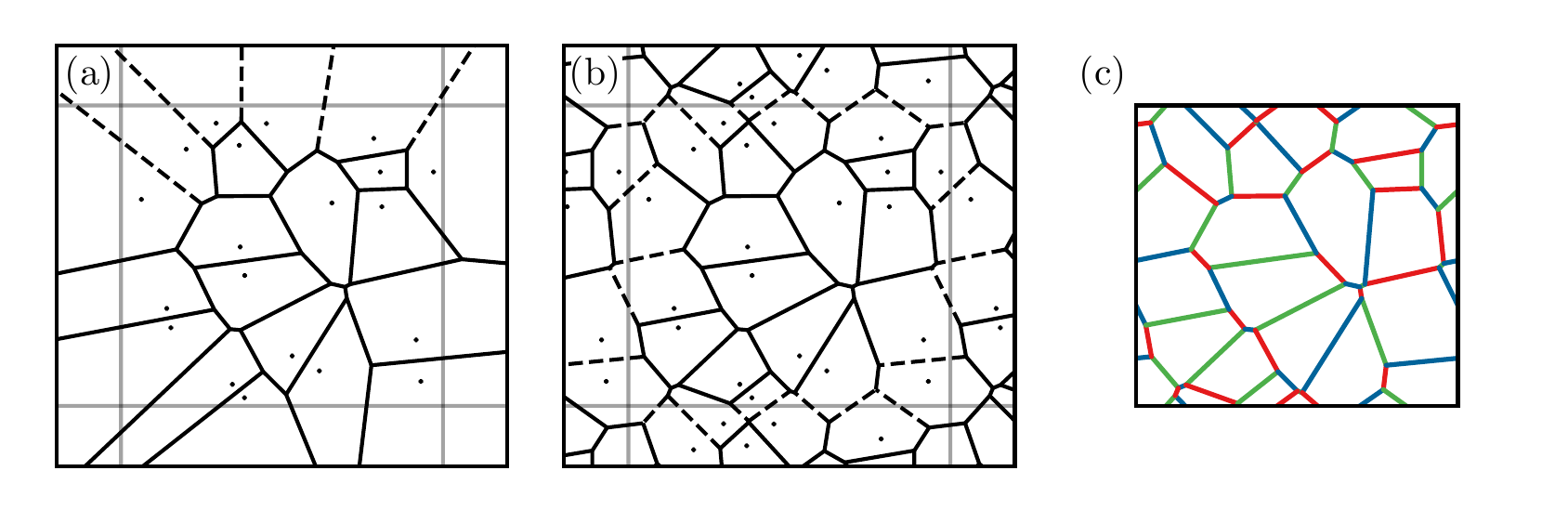}
    \caption{\textbf{(a)} The Voronoi partition (lines) splits a region up into polyhedra closer to one of the seed points (points) than any other. In two dimensions this yields a tri-coordinate lattice. Dotted lines go off to infinity. 
    \textbf{(b)} To create a lattice on the torus, we tile the seed points into a 3x3 grid and compute a Voronoi partition. By identifying pairs of edges (dotted lines) that cross the unit square (in grey) as the same we turn this lattice into on defined on the torus. 
    \textbf{(c)} The final tri-coordinate lattice in periodic boundary conditions, coloured such that all three colours meet at every vertex.}
    \label{fig:lattice_construction}
\end{figure*}

We generate tri-coordinate lattices by taking the Voronoi partition of the unit square with respect to uniformly sampled \textit{seed points}~\cite{florescu_designer_2009}. This partitions the space into polyhedral volumes enclosing the region closest to each seed point. In two dimensions, the vertices and edges of these polygons form a tri-coordinate lattice, exactly what is necessary for the Kitaev model.
To produce lattices with periodic boundary conditions we first tile the seed points into a repeating lattice. The Voronoi partition of the tiled seed points can then be converted into a lattice embedded onto the torus by connecting corresponding edges that cross the unit square boundaries, see~\cref{fig:lattice_construction}.
Finally, to somewhat regularise bond lengths in our lattice, a single step of Lloyd's algorithm is performed, where every vertex is shifted to the center of mass of the three plaquettes that it touches. This is done to improve the readability of the lattice, reducing the number of extremely short bonds appearing, and has no effect on the physics. 

Once a lattice has been generated, the bonds must be labelled in such a way that no vertex touches multiple edges of the same type, which we refer to as a \textit{three-edge colouring}. The problem of finding such a colouring is equivalent to the classical problem of four-colouring the faces, which is always solvable on a planar graph \cite{Tait1880, appelEveryPlanarMap1989a}. On the torus, a face colouring can require up to seven colours \cite{ringel_solution_1968}, so not all graphs can be assumed to be 3-edge colourable. However, such exceptions seem rare for graphs generated via voronisation -- every graph generated in this study admitted multiple distinct 3-edge colourings. In practice, the problem of finding a colouring for a given graph can be reduced to a Boolean satisfiability problem \cite{Karp1972}, which we then solve using the open-source solver \texttt{MiniSAT}~\cite{imms-sat18}.

Care must be taken in the definition of open boundary conditions, simply removing bonds from the lattice leaves behind unpaired \(b^\alpha\) operators that need to be paired in some way to arrive at fermionic modes. In order to fix a pairing we always start from a lattice defined on the torus and generate a lattice with open boundary conditions by defining the bond coupling \(J^{\alpha}_{ij} = 0\) for sites joined by bonds \((i,j)\) that we want to remove. This creates fermionic zero modes \(u_{ij}\) associated with these cut bonds which we set to 1 when calculating the projector. All our code is available online~\cite{koala}.

% \section{Euler Equation and Gauge Degeneracy}
% For a lattice with \(B\) bonds, \(V\) vertices, \(P\) plaquettes and genus \(g\) (1 for the torus) the Euler equation states that \(B = P + V + 2 - 2g\) which for the torus can be written as \(B = (P - 1) + (V - 1) + 2\). This corresponds nicely to the \(2^{B}\) gauge configurations being composed of \(2^{P - 1}\) physically distinct vortex states each of which is composed of \(2^{V - 1}\) gauge equivalent states that correspond to flipping three \(u_{ij}\) around a vertex. Finally there are 2 global flux operators that correspond to non-contractible loops on the torus.

\section{The Projector} \label{apx:projector}

Closely following the derivation of~\cite{pedrocchiPhysicalSolutionsKitaev2011} we can extend the projector from Majorana wavefunctions to physical spin states to the amorphous case. In the standard way, we define normal mode operators
\[(c_1, c_2... c_{2N}) Q = (b_1, b_1', b_2, b_2' ... b_{N}, b_{N}')\]
such that the Hamiltonian comes into the form
\[\tilde{H}_u = \frac{i}{2} \sum_m \epsilon_m b_m b_m'\]
from there we form fermionic operators \(f_i = \tfrac{1}{2} (b_m + ib_m')\)
and their associated number operators \(n_i = f^\dagger_i f_i\). The many body ground state within a vortex sector is then defined by the set of occupation numbers \(n_m = 0,1\). Lastly we need to define the fermion parity \(\hat{\pi} = \prod{i}^{N} (1 - 2\hat{n}_i)\).

The projector can be written as
\[ \mathcal{P} =  \mathcal{S} \left(\frac{1 + \prod_i^{2N} D_i}{2}\right) = \mathcal{S} \cdot \mathcal{P}_0\]
where \(D_i\) are the local projectors. \(\mathcal{S}\) symmetrises over gauge equivalent states while \(\mathcal{P}_0\) is responsible for annihilating unphysical states, see~\cite{pedrocchiPhysicalSolutionsKitaev2011} for details.

To extend this to the amorphous case we calculate the product of the local projectors \(D_i\)
\[\prod_i^{2N} D_i = \prod_i^{2N} b^x_i b^y_i b^z_i c_i \]
for a tri-coordinate lattice with \(N\) faces, \(2N\) vertices and \(3N\) edges. The operators can be ordered by bond type without utilising any property of the lattice.
\[\prod_i^{2N} D_i = \prod_i^{2N} b^x_i \prod_i^{2N} b^y_i \prod_i^{2N} b^z_i \prod_i^{2N} c_i\]
The product over \(c_i\) operators reduces to a determinant of the Q matrix and the fermion parity. The only problem is to compute the factors \(p_x,p_y,p_z = \pm1\) that arise from reordering the b operators such that pairs of vertices linked by the corresponding bonds are adjacent.
\[\prod_i^{2N} b^\alpha_i = p_\alpha \prod_{(i,j)}b^\alpha_i b^\alpha_j\]
This is simple the parity of the permutation from one ordering to the other and can be computed quickly with a cycle decomposition.

The final form is almost identical to the honeycomb case with the addition of the lattice structure factors \(p_x,p_y,p_z\)
\[P^0 = 1 + p_x\;p_y\;p_z \mathrm{det}(Q^u) \; \hat{\pi} \; \prod_{\{i,j\}} -iu_{ij},\] 

where \(\mathrm{det}(Q^u)\) and \(\prod u_{ij}\) depend on the lattice and the particular vortex sector. 

\section{Numerical Evidence for the Ground State Flux Sector} \label{apx:ground_state}

In this section we detail the numerical evidence collected to support the claim that, for an arbitrary lattice, a gapped ground state flux sector is found by setting the flux through each plaquette to $\phi_{\textup{g.s.}} = -(\pm i)^{n_{\textup{sides}}}$. This was done by generating a large number ($\sim$ 25,000) of lattices and exhaustively checking every possible flux sector to find the configuration with the lowest energy. We checked both the isotropic point ($J^\alpha = 1$), as well as in the toric code ($J^x = J^y = 0.25, J^z = 1$).\par
The argument has one complication: for a graph with $n_p$ plaquettes, there are $2^{n_p - 1}$ distinct flux sectors to search over, with an added factor of 4 when the global fluxes $\Phi_x$ and $\Phi_y$ wrapping around the cylinder directions are taken into account. Note that the $-1$ appears in this counting because fluxes can only be flipped in pairs. To be able to search over the entire flux space, one is necessarily restricted to looking at small system sizes -- we were able to check all flux sectors for systems with $n_p \leq 16$ in a reasonable amount of time. However, at such small system size we find that finite size effects are substantial. In order to overcome these effects we tile the system and use Bloch's theorem (a trick that we shall refer to as \textit{twist-averaging} for reasons that shall become clear) to efficiently find the energy of a much larger (but periodic) lattice. Thus we are able to suppress finite size effects, at the expense of losing long-range disorder in the lattice.\par
Our argument has three parts: First we shall detail the techniques used to exhaustively search the flux space for a given lattice. Next, we discuss finite-size effects and explain the way that our methods are modified by the twist-averaging procedure. Finally, we demonstrate that as the size of the disordered system is increased, the effect of twist-averaging becomes negligible -- suggesting that our conclusions still apply in the case of large disordered lattices. 

\begin{figure*}[t]
    \centering
    \includegraphics[width=0.9\textwidth]{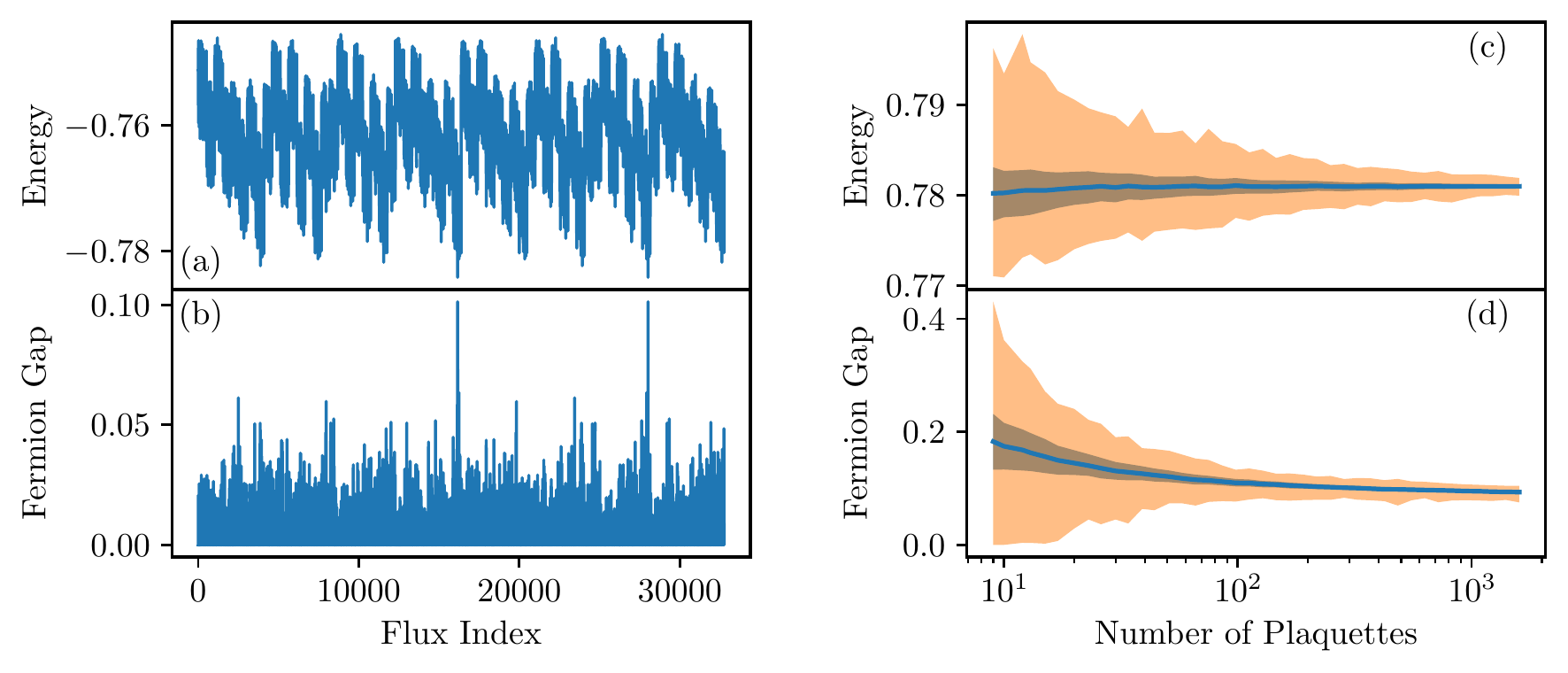}
    \caption{\textbf{(a)} The energy of every flux sector explored for a system of 16 plaquettes, the order is arbitrary. The two ground state flux sectors can be identified as the points with lowest energy. \textbf{(b)} The fermion gap for each of the flux sectors explored. Note that the largest fermion gap coincides with the ground state flux sector. This occurred in $\sim 85\%$ of cases tested. (c) Average energy of the systems tested over a range of system sizes from $n_p$ = 9 to $n_p = 1600$. The region between the upper and lower quartiles is shown in red, and the full range of energies obtained is shown in orange. (d) Average fermion gap as a function of system size. Again, the region between the upper and lower quartiles is shown in red, and the full range is shown in orange. As can be seen, no gapless systems were found for $n_p > 20$. }
    \label{fig:energy_gaps_example}
\end{figure*}

{\it Testing All Flux Sectors ---}
For a given lattice and flux sector, defined by $\{ u_{jk}\}$, the fermionic ground state energy is calculated by taking the sum of the negative eigenvalues of the matrix
\begin{align}
    M_{jk} = \frac{i}{2} J^{\alpha} u_{jk}.
\end{align}
The set of bond variables $u_{jk}$, which we are free to choose, determine the $\mathbb Z_2$ gauge field. However only the fluxes, defined for each plaquette according to 
\begin{equation}
    \phi_p = \prod_{(j,k) \in \partial p} -iu_{jk},
\end{equation}
have any effect on the energies. Thus, there is enormous degeneracy in the $u_{jk}$ degrees of freedom. Flipping the bonds along any closed loop on the dual lattice has no effect on the fluxes, since each plaquette has had an even number of its constituent bonds flipped - as is shown in the following diagram:
\begin{center}
    \begin{tikzpicture} 

\draw [line width=0.4mm,] (2.443089607848154 , 4.199174275000855) --  (2.133663274934363 , 3.699573812024347); 
\draw [line width=0.4mm,] (2.443089607848154 , 4.199174275000855) --  (2.826766606454033 , 4.129399589956555); 
\draw [line width=0.6mm,red!60] (2.826766606454033 , 4.129399589956555) --  (2.832350603254511 , 3.6412926004966697); 
\draw [line width=0.6mm,red!60] (2.133663274934363 , 3.699573812024347) --  (2.3533196431762136 , 3.3923797381398595); 
\draw [line width=0.4mm,] (2.832350603254511 , 3.6412926004966697) --  (2.3533196431762136 , 3.3923797381398595); 
\draw [line width=0.4mm,] (2.133663274934363 , 3.699573812024347) --  (1.6590601431591752 , 3.765943303309621); 
\draw [line width=0.4mm,] (1.7841101988472114 , 2.621425585623738) --  (1.3102001600764661 , 2.8828459026617206); 
\draw [line width=0.6mm,red!60] (1.7841101988472114 , 2.621425585623738) --  (2.2530706580695927 , 2.918572103800313); 
\draw [line width=0.4mm,] (1.3102001600764661 , 2.8828459026617206) --  (1.2704284082170167 , 3.4952254145963004); 
\draw [line width=0.4mm,] (2.2530706580695927 , 2.918572103800313) --  (2.3533196431762136 , 3.3923797381398595); 
\draw [line width=0.4mm,] (1.6590601431591752 , 3.765943303309621) --  (1.2704284082170167 , 3.4952254145963004); 
\draw [line width=0.4mm,] (2.832350603254511 , 3.6412926004966697) --  (3.2143675983672417 , 3.4321179123180476); 
\draw [line width=0.4mm,] (2.7336979624747713 , 2.772672717103165) --  (3.215963942694124 , 3.037305663264089); 
\draw [line width=0.4mm,] (2.7336979624747713 , 2.772672717103165) --  (2.2530706580695927 , 2.918572103800313); 
\draw [line width=0.4mm,] (3.215963942694124 , 3.037305663264089) --  (3.2143675983672417 , 3.4321179123180476); 
\draw [line width=0.4mm,] (1.66412589164998 , 4.27761376616966) --  (1.3628790737427836 , 4.554949458613537); 
\draw [line width=0.4mm,] (1.66412589164998 , 4.27761376616966) --  (2.1672741059313325 , 4.45048776187154); 
\draw [line width=0.4mm,] (1.6590601431591752 , 3.765943303309621) --  (1.66412589164998 , 4.27761376616966); 
\draw [line width=0.4mm,] (1.2704284082170167 , 3.4952254145963004) --  (0.7752117257338569 , 3.7981349822077344); 
\draw [line width=0.4mm,] (2.443089607848154 , 4.199174275000855) --  (2.1672741059313325 , 4.45048776187154); 
\draw [line width=0.4mm,] (3.603328735858143 , 4.2476394851768395) --  (3.5960004235960428 , 3.756977430689376); 
\draw [line width=0.4mm,] (3.603328735858143 , 4.2476394851768395) --  (4.039945897050822 , 4.297661569213179); 
\draw [line width=0.4mm,] (3.5960004235960428 , 3.756977430689376) --  (4.029745354756155 , 3.5751483526351198); 
\draw [line width=0.4mm,] (4.029745354756155 , 3.5751483526351198) --  (4.282495230506774 , 3.822901011217539); 
\draw [line width=0.4mm,] (4.282495230506774 , 3.822901011217539) --  (4.039945897050822 , 4.297661569213179); 
\draw [line width=0.4mm,] (2.3487551414914605 , 1.9800230663090006) --  (2.300540325954526 , 1.5282907169170128); 
\draw [line width=0.4mm,] (2.3487551414914605 , 1.9800230663090006) --  (1.8733514017504886 , 2.217815818486644); 
\draw [line width=0.4mm,] (1.8733514017504886 , 2.217815818486644) --  (1.602858071222523 , 1.785213933169); 
\draw [line width=0.4mm,] (1.602858071222523 , 1.785213933169) --  (1.8892256504947056 , 1.4056839700867723); 
\draw [line width=0.4mm,] (2.300540325954526 , 1.5282907169170128) --  (1.8892256504947056 , 1.4056839700867723); 
\draw [line width=0.4mm,] (4.668899029000971 , 1.370819996056893) --  (4.746438001722542 , 1.956213202683834); 
\draw [line width=0.4mm,] (4.746438001722542 , 1.956213202683834) --  (4.3482494616334675 , 2.151907119000129); 
\draw [line width=0.4mm,] (4.3482494616334675 , 2.151907119000129) --  (3.9290282747686036 , 1.6715546842085325); 
\draw [line width=0.4mm,] (4.177731391775351 , 1.2306194597430695) --  (3.9290282747686036 , 1.6715546842085325); 
\draw [line width=0.4mm,] (5.1341540722478065 , 2.0322396671565732) --  (4.746438001722542 , 1.956213202683834); 
\draw [line width=0.4mm,] (3.1199829882043235 , 1.5853535192856314) --  (3.5048132855156084 , 1.7811410414406073); 
\draw [line width=0.6mm,red!60] (3.1199829882043235 , 1.5853535192856314) --  (2.687310443097729 , 2.0354555882481775); 
\draw [line width=0.6mm,red!60] (3.5048132855156084 , 1.7811410414406073) --  (3.3721357795131874 , 2.2987943820561982); 
\draw [line width=0.4mm,] (3.3721357795131874 , 2.2987943820561982) --  (2.8464841372700485 , 2.4130019983886255); 
\draw [line width=0.4mm,] (2.687310443097729 , 2.0354555882481775) --  (2.8464841372700485 , 2.4130019983886255); 
\draw [line width=0.6mm,red!60] (2.3487551414914605 , 1.9800230663090006) --  (2.687310443097729 , 2.0354555882481775); 
\draw [line width=0.4mm,] (3.0121673012212615 , 1.1802036440941053) --  (3.1199829882043235 , 1.5853535192856314); 
\draw [line width=0.4mm,] (2.5312799405777326 , 1.1785733636646636) --  (2.300540325954526 , 1.5282907169170128); 
\draw [line width=0.4mm,] (1.8733514017504886 , 2.217815818486644) --  (1.7841101988472114 , 2.621425585623738); 
\draw [line width=0.4mm,] (2.7336979624747713 , 2.772672717103165) --  (2.8464841372700485 , 2.4130019983886255); 
\draw [line width=0.4mm,] (4.433307145184742 , 2.872234531196732) --  (4.063679393432637 , 3.1066171115286063); 
\draw [line width=0.4mm,] (4.433307145184742 , 2.872234531196732) --  (4.226815161490477 , 2.477492472211557); 
\draw [line width=0.6mm,red!60] (4.226815161490477 , 2.477492472211557) --  (3.674916468623193 , 2.514793378035552); 
\draw [line width=0.6mm,red!60] (4.063679393432637 , 3.1066171115286063) --  (3.6315308065994065 , 2.893633940528903); 
\draw [line width=0.4mm,] (3.6315308065994065 , 2.893633940528903) --  (3.674916468623193 , 2.514793378035552); 
\draw [line width=0.4mm,] (4.83894751221688 , 3.0786317478120138) --  (4.7220696362153936 , 3.5607669867263065); 
\draw [line width=0.4mm,] (4.83894751221688 , 3.0786317478120138) --  (4.433307145184742 , 2.872234531196732); 
\draw [line width=0.4mm,] (4.029745354756155 , 3.5751483526351198) --  (4.063679393432637 , 3.1066171115286063); 
\draw [line width=0.4mm,] (4.7220696362153936 , 3.5607669867263065) --  (4.282495230506774 , 3.822901011217539); 
\draw [line width=0.4mm,] (4.83894751221688 , 3.0786317478120138) --  (5.226527584560779 , 2.9637036802250862); 
\draw [line width=0.4mm,] (4.3482494616334675 , 2.151907119000129) --  (4.226815161490477 , 2.477492472211557); 
\draw [line width=0.4mm,] (3.9290282747686036 , 1.6715546842085325) --  (3.5048132855156084 , 1.7811410414406073); 
\draw [line width=0.4mm,] (3.3721357795131874 , 2.2987943820561982) --  (3.674916468623193 , 2.514793378035552); 
\draw [line width=0.4mm,] (3.215963942694124 , 3.037305663264089) --  (3.6315308065994065 , 2.893633940528903); 
\draw [line width=0.6mm,red!60] (3.2143675983672417 , 3.4321179123180476) --  (3.5960004235960428 , 3.756977430689376); 
\draw [line width=0.4mm,] (4.7220696362153936 , 3.5607669867263065) --  (4.901832297006263 , 3.8959627247133053); 
\draw [line width=0.4mm,] (1.3102001600764661 , 2.8828459026617206) --  (0.9205690962626919 , 2.7448693755016778); 
\draw [line width=0.4mm,] (1.602858071222523 , 1.785213933169) --  (1.1720730675383504 , 1.7504987593424772); 
\draw [line width=0.4mm,] (0.9205690962626919 , 2.7448693755016778) --  (0.8723030689515361 , 2.258574775850272); 
\draw [line width=0.4mm,] (0.8723030689515361 , 2.258574775850272) --  (1.1720730675383504 , 1.7504987593424772); 
\draw [line width=0.4mm,] (2.826766606454033 , 4.129399589956555) --  (3.015687026146436 , 4.486465145140302); 
\draw [line width=0.4mm,] (3.603328735858143 , 4.2476394851768395) --  (3.376487251015103 , 4.528709024461988); 
\draw [line width=0.4mm,] (0.9205690962626919 , 2.7448693755016778) --  (0.5128878465410877 , 3.0556589636233125); 
\draw [line width=0.4mm,] (0.8723030689515361 , 2.258574775850272) --  (0.46147593859581926 , 2.2588828762828306); 
\draw [line width=0.4mm,] (1.1720730675383504 , 1.7504987593424772) --  (0.9748432903949585 , 1.3652738897280694); 
\draw [blue!60,line width=0.4mm] plot [smooth cycle] coordinates { (2.0185904284584018, 2.769998844712026)  (2.2434914590552886, 3.5459767750821034)  (2.829558604854272, 3.8853460952266126)  (3.405184010981642, 3.5945476715037117)  (3.8476051000160214, 3.0001255260287545)  (3.950865815056835, 2.4961429251235545)  (3.438474532514398, 2.0399677117484027)  (2.903646715651026, 1.8104045537669045)  (2.5180327922945946, 2.007739327278589) };

\end{tikzpicture}
\end{center}
where the flipped bonds are shown in red. In order to explore every possible flux sector using the $u_{jk}$ variables, we restrict ourselves to change only a subset of the bonds in the system. In particular, we construct a spanning tree on the dual lattice, which passes through every plaquette in the system, but contains no loops. 
\begin{center}
    \begin{tikzpicture} 

\draw [line width=0.6mm, red!60] (2.443089607848154 , 4.199174275000855) --  (2.133663274934363 , 3.699573812024347); 
\draw [line width=0.4mm, blue!60] (2.5178379471334553 , 3.812364003123657) --  (2.013442604704601 , 4.078558583675204); 
\draw [line width=0.4mm,] (2.443089607848154 , 4.199174275000855) --  (2.826766606454033 , 4.129399589956555); 
\draw [line width=0.4mm,] (2.826766606454033 , 4.129399589956555) --  (2.832350603254511 , 3.6412926004966697); 
\draw [line width=0.4mm,] (2.133663274934363 , 3.699573812024347) --  (2.3533196431762136 , 3.3923797381398595); 
\draw [line width=0.6mm, red!60] (2.832350603254511 , 3.6412926004966697) --  (2.3533196431762136 , 3.3923797381398595); 
\draw [line width=0.4mm, blue!60] (2.767128401339409 , 3.1990567891870243) --  (2.5178379471334553 , 3.812364003123657); 
\draw [line width=0.6mm, red!60] (2.133663274934363 , 3.699573812024347) --  (1.6590601431591752 , 3.765943303309621); 
\draw [line width=0.4mm, blue!60] (1.8234074980685773 , 3.2537094085937) --  (2.013442604704601 , 4.078558583675204); 
\draw [line width=0.6mm, red!60] (1.7841101988472114 , 2.621425585623738) --  (1.3102001600764661 , 2.8828459026617206); 
\draw [line width=0.4mm, blue!60] (1.3622092949498954 , 2.3230348786622184) --  (1.8234074980685773 , 3.2537094085937); 
\draw [line width=0.4mm,] (1.7841101988472114 , 2.621425585623738) --  (2.2530706580695927 , 2.918572103800313); 
\draw [line width=0.4mm,] (1.3102001600764661 , 2.8828459026617206) --  (1.2704284082170167 , 3.4952254145963004); 
\draw [line width=0.4mm,] (2.2530706580695927 , 2.918572103800313) --  (2.3533196431762136 , 3.3923797381398595); 
\draw [line width=0.6mm, red!60] (1.6590601431591752 , 3.765943303309621) --  (1.2704284082170167 , 3.4952254145963004); 
\draw [line width=0.4mm, blue!60] (1.8234074980685773 , 3.2537094085937) --  (1.2047281385030177 , 4.086249239531997); 
\draw [line width=0.4mm,] (2.832350603254511 , 3.6412926004966697) --  (3.2143675983672417 , 3.4321179123180476); 
\draw [line width=0.6mm, red!60] (2.7336979624747713 , 2.772672717103165) --  (3.215963942694124 , 3.037305663264089); 
\draw [line width=0.4mm, blue!60] (2.767128401339409 , 3.1990567891870243) --  (3.245788182862455 , 2.6550336798960887); 
\draw [line width=0.4mm,] (2.7336979624747713 , 2.772672717103165) --  (2.2530706580695927 , 2.918572103800313); 
\draw [line width=0.6mm, red!60] (3.215963942694124 , 3.037305663264089) --  (3.2143675983672417 , 3.4321179123180476); 
\draw [line width=0.4mm, blue!60] (2.767128401339409 , 3.1990567891870243) --  (3.6252145865742675 , 3.300300068494024); 
\draw [line width=0.4mm,] (1.66412589164998 , 4.27761376616966) --  (1.3628790737427836 , 4.554949458613537); 
\draw [line width=0.6mm, red!60] (1.66412589164998 , 4.27761376616966) --  (2.1672741059313325 , 4.45048776187154); 
\draw [line width=0.4mm, blue!60] (1.7982635613900646 , 4.692115720838544) --  (2.013442604704601 , 4.078558583675204); 
\draw [line width=0.4mm,] (1.6590601431591752 , 3.765943303309621) --  (1.66412589164998 , 4.27761376616966); 
\draw [line width=0.4mm,] (1.2704284082170167 , 3.4952254145963004) --  (0.7752117257338569 , 3.7981349822077344); 
\draw [line width=0.4mm,] (2.443089607848154 , 4.199174275000855) --  (2.1672741059313325 , 4.45048776187154); 
\draw [line width=0.6mm, red!60] (3.603328735858143 , 4.2476394851768395) --  (3.5960004235960428 , 3.756977430689376); 
\draw [line width=0.4mm, blue!60] (3.910303128353587 , 3.9400655697864106) --  (3.20928403495593 , 4.03180016974854); 
\draw [line width=0.4mm,] (3.603328735858143 , 4.2476394851768395) --  (4.039945897050822 , 4.297661569213179); 
\draw [line width=0.6mm, red!60] (3.5960004235960428 , 3.756977430689376) --  (4.029745354756155 , 3.5751483526351198); 
\draw [line width=0.4mm, blue!60] (3.910303128353587 , 3.9400655697864106) --  (3.6252145865742675 , 3.300300068494024); 
\draw [line width=0.4mm,] (4.029745354756155 , 3.5751483526351198) --  (4.282495230506774 , 3.822901011217539); 
\draw [line width=0.6mm, red!60] (4.282495230506774 , 3.822901011217539) --  (4.039945897050822 , 4.297661569213179); 
\draw [line width=0.4mm, blue!60] (3.910303128353587 , 3.9400655697864106) --  (4.474016274392756 , 4.104509158153612); 
\draw [line width=0.4mm,] (2.3487551414914605 , 1.9800230663090006) --  (2.300540325954526 , 1.5282907169170128); 
\draw [line width=0.6mm, red!60] (2.3487551414914605 , 1.9800230663090006) --  (1.8733514017504886 , 2.217815818486644); 
\draw [line width=0.4mm, blue!60] (2.002946118182741 , 1.7834055009936858) --  (2.3609685632859003 , 2.422709553994238); 
\draw [line width=0.4mm,] (1.8733514017504886 , 2.217815818486644) --  (1.602858071222523 , 1.785213933169); 
\draw [line width=0.4mm,] (1.602858071222523 , 1.785213933169) --  (1.8892256504947056 , 1.4056839700867723); 
\draw [line width=0.6mm, red!60] (2.300540325954526 , 1.5282907169170128) --  (1.8892256504947056 , 1.4056839700867723); 
\draw [line width=0.4mm, blue!60] (2.163240614722578 , 1.1359652989758509) --  (2.002946118182741 , 1.7834055009936858); 
\draw [line width=0.6mm, red!60] (4.668899029000971 , 1.370819996056893) --  (4.746438001722542 , 1.956213202683834); 
\draw [line width=0.4mm, blue!60] (4.3740692317801875 , 1.6762228923384912) --  (5.0359216553494415 , 1.5406491898288264); 
\draw [line width=0.4mm,] (4.746438001722542 , 1.956213202683834) --  (4.3482494616334675 , 2.151907119000129); 
\draw [line width=0.6mm, red!60] (4.3482494616334675 , 2.151907119000129) --  (3.9290282747686036 , 1.6715546842085325); 
\draw [line width=0.4mm, blue!60] (4.3740692317801875 , 1.6762228923384912) --  (3.8426597385907564 , 2.1492805128254298); 
\draw [line width=0.4mm,] (4.177731391775351 , 1.2306194597430695) --  (3.9290282747686036 , 1.6715546842085325); 
\draw [line width=0.4mm,] (5.1341540722478065 , 2.0322396671565732) --  (4.746438001722542 , 1.956213202683834); 
\draw [line width=0.4mm,] (3.1199829882043235 , 1.5853535192856314) --  (3.5048132855156084 , 1.7811410414406073); 
\draw [line width=0.6mm, red!60] (3.1199829882043235 , 1.5853535192856314) --  (2.687310443097729 , 2.0354555882481775); 
\draw [line width=0.4mm, blue!60] (2.666672690091172 , 1.5813166497530986) --  (3.106145326720179 , 2.022749305883848); 
\draw [line width=0.4mm,] (3.5048132855156084 , 1.7811410414406073) --  (3.3721357795131874 , 2.2987943820561982); 
\draw [line width=0.6mm, red!60] (3.3721357795131874 , 2.2987943820561982) --  (2.8464841372700485 , 2.4130019983886255); 
\draw [line width=0.4mm, blue!60] (3.106145326720179 , 2.022749305883848) --  (3.245788182862455 , 2.6550336798960887); 
\draw [line width=0.4mm,] (2.687310443097729 , 2.0354555882481775) --  (2.8464841372700485 , 2.4130019983886255); 
\draw [line width=0.4mm,] (2.3487551414914605 , 1.9800230663090006) --  (2.687310443097729 , 2.0354555882481775); 
\draw [line width=0.4mm,] (3.0121673012212615 , 1.1802036440941053) --  (3.1199829882043235 , 1.5853535192856314); 
\draw [line width=0.6mm, red!60] (2.5312799405777326 , 1.1785733636646636) --  (2.300540325954526 , 1.5282907169170128); 
\draw [line width=0.4mm, blue!60] (2.163240614722578 , 1.1359652989758509) --  (2.666672690091172 , 1.5813166497530986); 
\draw [line width=0.4mm,] (1.8733514017504886 , 2.217815818486644) --  (1.7841101988472114 , 2.621425585623738); 
\draw [line width=0.4mm,] (2.7336979624747713 , 2.772672717103165) --  (2.8464841372700485 , 2.4130019983886255); 
\draw [line width=0.6mm, red!60] (4.433307145184742 , 2.872234531196732) --  (4.063679393432637 , 3.1066171115286063); 
\draw [line width=0.4mm, blue!60] (4.006049795066091 , 2.7729542867002706) --  (4.395040712052096 , 3.3360499568527193); 
\draw [line width=0.4mm,] (4.433307145184742 , 2.872234531196732) --  (4.226815161490477 , 2.477492472211557); 
\draw [line width=0.6mm, red!60] (4.226815161490477 , 2.477492472211557) --  (3.674916468623193 , 2.514793378035552); 
\draw [line width=0.4mm, blue!60] (3.8426597385907564 , 2.1492805128254298) --  (4.006049795066091 , 2.7729542867002706); 
\draw [line width=0.6mm, red!60] (4.063679393432637 , 3.1066171115286063) --  (3.6315308065994065 , 2.893633940528903); 
\draw [line width=0.4mm, blue!60] (4.006049795066091 , 2.7729542867002706) --  (3.6252145865742675 , 3.300300068494024); 
\draw [line width=0.4mm,] (3.6315308065994065 , 2.893633940528903) --  (3.674916468623193 , 2.514793378035552); 
\draw [line width=0.4mm,] (4.83894751221688 , 3.0786317478120138) --  (4.7220696362153936 , 3.5607669867263065); 
\draw [line width=0.6mm, red!60] (4.83894751221688 , 3.0786317478120138) --  (4.433307145184742 , 2.872234531196732); 
\draw [line width=0.4mm, blue!60] (4.861599986375169 , 2.4896454558683327) --  (4.395040712052096 , 3.3360499568527193); 
\draw [line width=0.4mm,] (4.029745354756155 , 3.5751483526351198) --  (4.063679393432637 , 3.1066171115286063); 
\draw [line width=0.4mm,] (4.7220696362153936 , 3.5607669867263065) --  (4.282495230506774 , 3.822901011217539); 
\draw [line width=0.4mm,] (4.83894751221688 , 3.0786317478120138) --  (5.226527584560779 , 2.9637036802250862); 
\draw [line width=0.4mm,] (4.3482494616334675 , 2.151907119000129) --  (4.226815161490477 , 2.477492472211557); 
\draw [line width=0.4mm,] (3.9290282747686036 , 1.6715546842085325) --  (3.5048132855156084 , 1.7811410414406073); 
\draw [line width=0.4mm,] (3.3721357795131874 , 2.2987943820561982) --  (3.674916468623193 , 2.514793378035552); 
\draw [line width=0.4mm,] (3.215963942694124 , 3.037305663264089) --  (3.6315308065994065 , 2.893633940528903); 
\draw [line width=0.4mm,] (3.2143675983672417 , 3.4321179123180476) --  (3.5960004235960428 , 3.756977430689376); 
\draw [line width=0.4mm,] (4.7220696362153936 , 3.5607669867263065) --  (4.901832297006263 , 3.8959627247133053); 
\draw [line width=0.6mm, red!60] (1.3102001600764661 , 2.8828459026617206) --  (0.9205690962626919 , 2.7448693755016778); 
\draw [line width=0.4mm, blue!60] (1.3622092949498954 , 2.3230348786622184) --  (0.8660999107838037 , 3.245546616164256); 
\draw [line width=0.4mm,] (1.602858071222523 , 1.785213933169) --  (1.1720730675383504 , 1.7504987593424772); 
\draw [line width=0.4mm,] (0.9205690962626919 , 2.7448693755016778) --  (0.8723030689515361 , 2.258574775850272); 
\draw [line width=0.6mm, red!60] (0.8723030689515361 , 2.258574775850272) --  (1.1720730675383504 , 1.7504987593424772); 
\draw [line width=0.4mm, blue!60] (1.3622092949498954 , 2.3230348786622184) --  (0.7221741160928734 , 1.83704393836023); 
\draw [line width=0.4mm,] (2.826766606454033 , 4.129399589956555) --  (3.015687026146436 , 4.486465145140302); 
\draw [line width=0.4mm,] (3.603328735858143 , 4.2476394851768395) --  (3.376487251015103 , 4.528709024461988); 
\draw [line width=0.6mm, red!60] (0.9205690962626919 , 2.7448693755016778) --  (0.5128878465410877 , 3.0556589636233125); 
\draw [line width=0.4mm, blue!60] (0.5279349045632474 , 2.654532424478553) --  (0.8660999107838037 , 3.245546616164256); 
\draw [line width=0.4mm,] (0.8723030689515361 , 2.258574775850272) --  (0.46147593859581926 , 2.2588828762828306); 
\draw [line width=0.4mm,] (1.1720730675383504 , 1.7504987593424772) --  (0.9748432903949585 , 1.3652738897280694); 

\end{tikzpicture}

\end{center}
The tree contains $n_p - 1$ edges, shown in red, whose configuration space has a $1:1$ mapping onto the $2^{n_p - 1}$ distinct flux sectors. Each flux sector can be created in precisely one way by flipping edges only on the tree (provided all other bond variables not on the tree remain fixed). Thus, all possible flux sectors can be accessed by iterating over all configurations of edges on this spanning tree.

{\it Finite Size Effects ---}
In our numerical investigation, the objective was to test as many example lattices as possible. We aim for the largest lattice size that could be efficiently solved, requiring a balance between lattice size and cases tested. Each added plaquette doubles the number of flux sectors that must be checked. 25,000 lattices containing 16 plaquettes were used. However, in his numerical investigation of the honeycomb model, Kitaev demonstrated that finite size effects persist up to much larger lattice sizes than we were able to access \cite{kitaevAnyonsExactlySolved2006}. \par 
In order to circumvent this problem, we treat the 16-plaquette amorphous lattice as a unit cell in an arbitrarily large periodic system. The bonds that originally connected across the periodic boundaries now connect adjacent unit cells. This infinite periodic Hamiltonian can then be solved using Bloch's theorem, since the larger system is diagonalised by a plane wave ansatz. For a given crystal momentum $\bf q \in [0,2\pi)^2 $, we are left with a Bloch Hamiltonian, which is identical to the original Hamiltonian aside from an extra phase on edges that cross the periodic boundaries in the $x$ and $y$ directions,
\begin{align}
    M_{jk}(\bf q) =  \frac{i}{2} J^{\alpha} u_{jk} e^{i q_{jk}},
\end{align}
where $q_{jk} = q_x$ for a bond that crosses the $x$-periodic boundary in the positive direction, with the analogous definition for $y$-crossing bonds. We also have $q_{jk} = -q_{kj}$. Finally $q_{jk} = 0$ if the edge does not cross any boundaries at all -- in essence we are imposing twisted boundary conditions on our system. The total energy of the tiled system can be calculated by summing the energy of $M(\bf q)$ for every value of $\bf q$. In practice we constructed a lattice of $50 \times 50$ values of $\bf q$ spanning the Brillouin zone. The procedure is called twist averaging because the energy-per-unit cell is equivalent to the average energy over the full range of twisted boundary conditions. \par 
{\it Evidence for the Ground State Ansatz ---}
For each lattice with 16 plaquettes, $2^{15} =$ 32,768 flux sectors are generated. In each case we find the energy (averaged over all twist values) and the size of the fermion gap, which is defined as the lowest energy excitation for any value of $\bf q$. We then check if the lowest energy flux sector aligns with our ansatz (given by $\phi^{\textup{g.s.}}_p = -(\pm i)^{n_{\textup{sides}}}$) and whether this flux sector is gapped. \par
In the isotropic case ($J^\alpha = 1$), all 25,000 examples conformed to our guess for the ground state flux sector. A tiny minority ($\sim 10$) of the systems were found to be gapless. As we shall see shortly, the proportion of gapless systems vanishes as we increase the size of the amorphous lattice. An example of the energies and gaps for one of the systems tested is shown in fig.~\ref{fig:energy_gaps_example} (a). For the anisotropic phase (we used $ J^x, J^y = 0.25, J^z = 1$) the overwhelming majority of cases adhered to our ansatz, however a small minority ($\sim 0.5 \%$) did not. In these cases, however, the energy difference between our ansatz and the ground state was at most of order $10^{-6}$. Further investigation would need to be undertaken to determine whether these anomalous systems are a finite size effect due to the small amorphous system sizes used or a genuine feature of the toric code phase on such lattices. \par

{\it A Gapped Ground State ---} Now that we have collected sufficient evidence to support our guess for the ground state flux sector, we turn our attention to checking that this sector is gapped. We no longer need to exhaustively search over flux space for the ground state, so it is possible to go to much larger system size. We generate 40 sets of systems with plaquette numbers ranging from 9 to 1600. For each system size, 1000 distinct lattices are generated and the energy and gap size are calculated without phase twisting, since the effect is negligible for such large system sizes. As can be seen in fig.~\ref{fig:energy_gaps_example} (c-d), for very small system size a small minority of gapless systems appear, however beyond around 20 plaquettes all systems had a stable fermion gap in the ground state. Finally, the energy and gap difference between the phase twisted and non-phase twisted results vanished exponentially with the system size, supporting the claim that the results can be straightforwardly extrapolated to large systems.

\section{The Effect of Bond Disorder}
\label{apx:bond-dis}

\begin{figure*}
    \centering
    \includegraphics[width=0.9\textwidth]{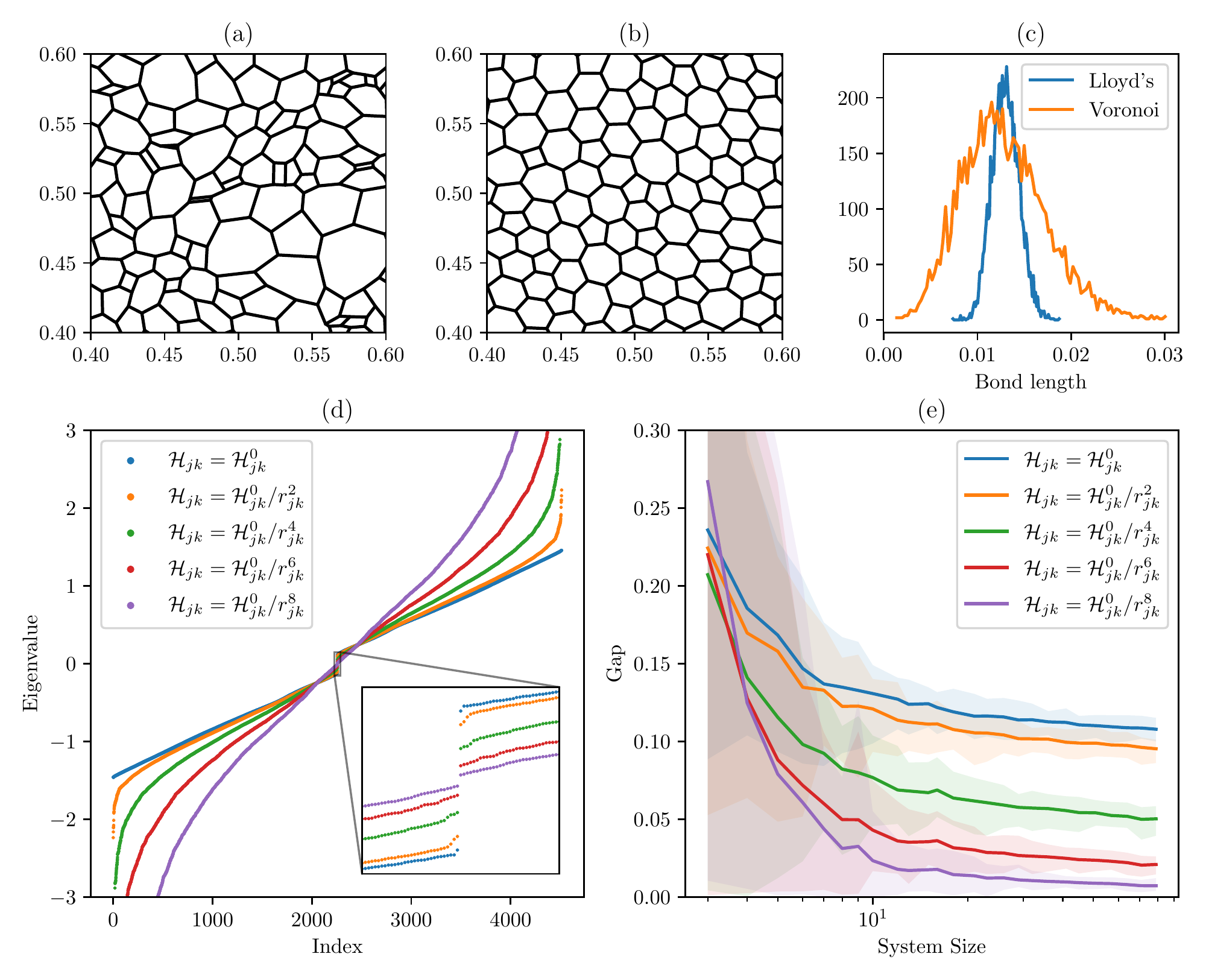}
    \caption{
    \textbf{(a)} Close up of a section of amorphous lattice generated using a single voronoi partition on a set of uniformly sampled seed points. \textbf{(b)} A section of amorphous lattice generated using the regularised method detailed in appendix \ref{apx:bond-dis}, by creating a Voronoi partition using seed points sampled from a blue noise distribution. This is then followed by four iterations of Lloyd's algorithm to further regularise the bond lengths. \textbf{(c)} A histogram of bond lengths for the example graphs in sub figures a and b. Note that the distribution is much more tightly peaked around the mean bond length in the regularised lattice, with no extremely short bonds present. \textbf{(d)} The Majorana spectrum for a single example of an amorphous lattice for five different values of $\alpha \in  [0 , 8]$. \textbf{(e)} Finite-size scaling analysis of the fermion gap in the flux-free sector. Solid lines indicate the gap averaged over 50 random lattice samples for each system size, and the shaded regions indicate the full range of the sampled gaps.
    } 
    \label{fig:disorder}
\end{figure*}

The effect of bond disorder in the Kitaev honeycomb lattice model is well studied~\cite{knolle_dynamics_2016, knolleBondDisorderedSpinLiquid2019, selfThermallyInducedMetallic2019}. When bond disorder is introduced the fermionic gap closes as a function of disorder strength, leading generically to a thermal metal phase~\cite{selfThermallyInducedMetallic2019, knolle_dynamics_2016}. The changes to the fermionic spectrum of the flux-free sector are much more dramatic than those observed in our amorphous model. \emph{A priori} one might expect that a similar gap closing should be introduced by the lattice disorder present in the amorphous model. As we have shown, this is not the case. We believe the essential difference between these kinds of disorder is the disruption of the local motif: in the amorphous model, the global translational symmetry of the system is disrupted, but the coupling strengths remain locally homogeneous on each site.

We study the effect of bond disorder on the amorphous lattice by introducing a bond-length dependent factor to the hopping terms in the Hamiltonian, such that $J_{jk} = J_{jk}^0 / r_{jk}^\alpha$, where $r_{jk}$ is the bond length from site $j$ to site $k$, which has been normalised such that $\langle r_{jk}^\alpha \rangle = 1$ to prevent this term from substantially rescaling the energies of the system. The parameter $\alpha$ controls the strength of the disorder, where setting $\alpha=0$ reproduces the original Hamiltonian. As $\alpha$ is varied, we calculate the fermionic spectrum, observing how the band gap is affected by this disorder. 

Clearly, this Hamiltonian will produce unphysical results on any lattice where some bonds are extremely short-ranged, leading to extremely strong hopping terms. Thus, in order to get physical results, one must first regularise the bond lengths of the lattice. This is done in two steps. First, the initial points are sampled from a blue noise distribution, rather than a uniform distribution, using Mitchell's best-candidate algorithm \cite{mitchell_1991, demofox2_generating_2017}. This produces seed points that are regularly spaced, yet still amorphous. The second step is to regularise the lattice using Lloyd's algorithm \cite{lloyd_1982} -- where vertices are repeatedly moved to the centre of mass of the plaquettes adjacent and the Voronoi diagram is regenerated. In practice we used four iterations of this method. An example of such a lattice is shown in fig.~\ref{fig:disorder} (b).\par 

The fermionic spectrum is calculated for a range of $\alpha$ between 0 and 8, at a variety of lattice sizes parametrised by the length $L$, where the number of vertices $N$ scales with $N \sim L^2 $. As disorder is increased, the gap shrinks however no truly gapless systems are generated for systems sizes larger than $L\sim 10$, providing evidence of the Kitaev phase's stability in the presence of even relatively strong bond disorder.We note that these calculations were carried out in periodic boundary conditions, hence the absence of in-gap edge states.

\section{Evidence for an Anderson Transition to a Thermal Metal Phase} \label{apx:thermal_metal}

\begin{figure*}
    \centering
    \includegraphics[width=0.95\textwidth]{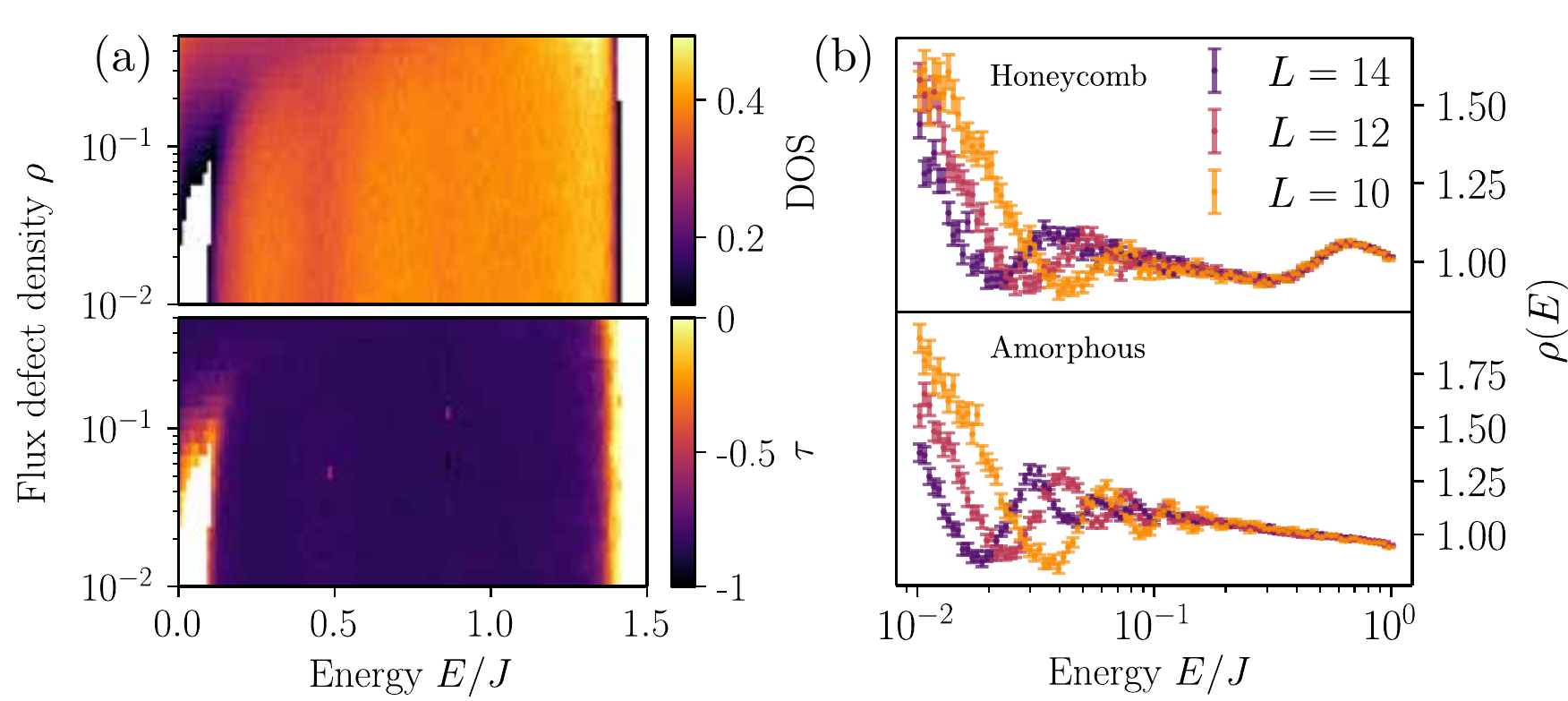}
    \caption{\textbf{(a)} Density of states (top) and inverse participation ratio scaling exponent (bottom) of the fermionic spectrum as a function of flux defect density, $\rho$, for isotropic couplings. Each pixel is averaged over 10 independent lattice realisations, in flux sectors sampled from an ensemble with a proportion $\rho$ of fluxes flipped with respect to the ground state sector. White pixels correspond to bins containing no fermionic states. At low defect density, the fermionic spectra are gapped. As the defect density increases, in-gap states appear with a small $\tau$, indicating that they are strongly localized around defects. At large defect density, $\tau$ increases for the in-gap sites, indicating that they are delocalised and the system becomes gapless. Using defect density as a proxy for temperature, this demonstrates the thermal phase transition from a chiral spin liquid to a thermal metal phase. 
    \textbf{(b)} A histogram of fermionic density of states sampled from the thermodynamic ensemble of flux sectors for $T\to\infty$, i.e.~all gauge configurations equally likely. The oscillations at low $E$ are characteristic of a thermal metal phase~\cite{selfThermallyInducedMetallic2019}, demonstrated for the Kitaev honeycomb lattice model subject to a magnetic field (top) and the amorphous Kitaev model (bottom). $L$ corresponds to the linear extent of the system -- $L\sim\sqrt{N}$ with $N$ sites -- for both lattice types.}
    \label{fig:DOS_Oscillations}
\end{figure*}

Here we present numerical evidence to support the claim that, as temperature is increased, the gapped chiral spin liquid undergoes an Anderson transition to a gapless thermal metal phase \cite{Laumann2012, selfThermallyInducedMetallic2019, Chalker_thermal_2001}. As discussed in the bulk text, we look for two signatures of this transition: a closing of the fermion gap driven by the flux sectors with a gapless fermionic spectrum, and the characteristic low energy oscillations in the density of states predicted by random matrix theory (RMT) \cite{bocquet_disordered_2000, selfThermallyInducedMetallic2019}.\par 
We study the closing of the fermion gap using the flux density $\rho$ as a proxy for temperature. This approximation is exact in the limits $T = 0$ ($\rho = 0$) and $T \to \infty$ ($\rho = 0.5$). At intermediate temperatures the method neglects the influence of flux-flux correlations. However, we are only interested in whether the gap closes at all. The fermionic density of states as a function of $\rho$ is shown in~\cref{fig:DOS_Oscillations}(a). As the defect density increases, the gap becomes populated with fermionic states. We quantify the degree to which a state is localised by calculating the dimensional scaling exponent of the IPR with the linear extent of the system, $L\sim\sqrt{N}$, with $N$ being the number of sites on the lattice,
\begin{equation}
    \mathrm{IPR} \propto L^{-\tau}.
\end{equation}
At small $\rho$, the states populating the gap possess $\tau\approx0$, indicating that they are localised states pinned to individual fluxes -- the system remains insulating. At larger $\rho$, the in-gap states merge with the bulk band and become extensive, closing the gap -- the system transitions to a metallic phase.

Finally, the averaged density of states in the $\rho = 0.5$ case is shown in \cref{fig:DOS_Oscillations} (b) for both the Honeycomb model and our amorphous lattice. Note that only the honeycomb model is calculated in the presence of an effective magnetic field explicitly breaking TRS. In both cases we see the logarithmic scaling alongside the characteristic RMT oscillations at low energy, giving strong evidence that the amorphous model displays a finite temperature transition to a thermal metal phase.

\end{document}